\documentclass[aps,pra,amsmath,amssymb,amsfonts,superscriptaddress,floatfix,nofootinbib]{revtex4} 
\pdfoutput=1
\usepackage{amsmath,amsthm,amsfonts,amssymb}
\usepackage{graphicx}

\usepackage{color}
\usepackage{hyperref}

\usepackage{algorithm}
\usepackage[capitalize,nameinlink]{cleveref}

\newcommand{\be}{\begin{equation}}
\newcommand{\ee}{\end{equation}}

\begin{document}

\title{A Class of Cyclic Quantum Codes}
\author{Matthew B.~Hastings}
\begin{abstract}
We introduce a class of cyclic quantum codes, basing the construction not on the simplicity of the stabilizers, but rather on the simplicity of preparation of a code state (at least in the absence of noise).  We show how certain known codes, such as a certain family of rotated two-dimensional toric codes, fall into this class, and we also give certain other examples at small sizes found by computer search.  We finally discuss fault tolerant preparation of these codes.
\end{abstract}
\maketitle

We introduce a class of cyclic quantum codes defined using a bipartite cluster state.
After a Hadamard transformation on certain qubits, these are CSS quantum codes.
We term these codes BCC codes (``bipartite cyclic cluster" codes).

Generally, for quantum stabilizer codes, there is some interest in finding ``simpler" codes.  This has often been interpreted as wanting the stabilizers to be low weight, as this simplifies the circuits to measure the syndrome.  Here, we consider a different notion of a simpler code, asking for simplified circuits to prepare a code state initialized with a particular value of logicals.  We give a general definition of these codes in \cref{gd} and specific properties, followed by showing that certain known codes fall into this class in \cref{rtc}, and giving some small examples found by computer search in \cref{oe}.
For the majority of this, we will consider preparation circuits in the absence of noise, only considering fault tolerant preparation in \cref{ftp}.

The family of codes considered here has some advantages.  They may provide simpler preparation circuits.  They are rich enough to exhibit reasonable distance, and yet they are structured enough that they are amenable to computer search.  Their structure may  be useful in implementations in neutral atom systems where the atoms are moved near each other before implementing gates between them, as this structure may simplify the moves required. Some of the BCC codes are a special case of bicycle codes in \cite{kovalev2013quantum}; the interest here is the emphasis on the simpler preparation circuit, meaning that some codes with relatively high weight stabilizers (and so being relatively complicated as bicycle codes) are simple from this point of view.

\section{General Definition}
\label{gd}
In general, we will consider $[[n,k,d]]$ codes where $n=k n_0$ for some given $n_0$.
We label the qubits by a pair $(m,j)$ where $m$ is defined mod $n_0$ and where $1\leq j\leq k$.
For a given code, define a basis of code states as follows; these basis states are those which can be prepared by, for each $j$, initializing all qubits $(m,j)$ in either the $+$ or the $-$ state, and then applying a certain unitary $U$, so that there are $2^k$ basis states.
We require that the unitary $U$ be a product of CZ gates obeying certain properties below.
Because the state is created by initializing qubits to $+$ (or $-$) and applying a product of CZ gates, the resulting state is a cluster state\cite{briegel2001persistent} (or possibly a cluster state up to applying Pauli $Z$ to certain qubits which are initialized in the $-$ state).

Such a cluster state can be described by a graph $G$: vertices correspond to qubits and there is an edge between vertices if a CZ gates acts on those two qubits (we assume that for each pair of qubits, there are either zero or one CZ gates acting on that pair).
To state the properties of $U$ in terms of this graph $G$, this graph should be bipartite and have a certain cyclic invariance.

Formally:
to define bipartiteness, we define some set ${\cal A}\subset \{1,\ldots,k\}$ and we define the set ${\cal B}$ to be $\{1,\ldots,k\}\setminus {\cal A}$.  We define two sets of qubits $A,B$, where $A$ is the set of qubits $(m,j)$ with $j\in{\cal A}$ and $B$ is the set of qubits $(m,j)$ with $j\in {\cal B}$.
Then we require that
{\bf (1):} each CZ gate 
acts on one qubit in $A$ and one qubit in $B$, and {\bf (2):} $U$ is cyclically invariant, in that if a gate acts on qubits $(m,j)$ and $(m',j')$ then for each $l$, there is a gate acting on $(m+l,j)$ and $(m'+l,j')$.  

Thus, the parameters of the code are $n_0,k$, and the given graph $G$.
This code indeed has the given number of logical qubits $k$ as for each $j\in \{1,\ldots,k\}$ we may choose to either initialize in $+$ or in $-$.

By applying a Hadamard on qubits in $B$ to the resulting code state, the code is a CSS quantum code.  The resulting code can be described by initializing qubits in $A$ in $+$ or $-$ for each $j$, initializing qubits in $B$ in $0$ or $1$ for each $j$, calling the resulting state the ``initial state",and then applying CNOT gates with source in $A$ and target in $B$, calling the resulting state the ``code state".
{\emph For the rest of this paper, we will use this CSS quantum code.}
Note that the initial state is a repetition code in the $+,-$ basis for qubits in $A$ and in the $0,1$ basis for qubits in $B$.

We may compute the stabilizers of the code as follows.
For each $j\in {\cal A}$, for each $m$, the operator $X_{m,j} X_{m+1,j}$ is a stabilizer of the initial state.  Hence, $U X_{m_j} X_{m+1,j} U^\dagger$ is a stabilizer of the code state.  This operator is
 $X_{m,j} X_{m+1,j}$ times a product of $X$ over qubits in $B$ whose corresponding vertices share an odd number of edges with $(m,j)$ or $(m+1,j)$.
Similarly, for each $j \in {\cal B}$, the operators $Z_{m,j} Z_{m+1,j}$ is a stabilizer of the initial state and therefore a stabilizer of the code state is
$Z_{m,j} Z_{m+1,j}$ times a product of $Z$ over qubits in $B$ whose corresponding vertices share an odd number of edges with $(m,j)$ or $(m+1,j)$.

From the stabilizers of the code, we may see that these BCC codes are a special case of generalized bicycle codes in \cite{kovalev2013quantum}, with a choice of $A$ having weight two.  The particular interest here is the simple class of circuit giving these codes.

We may easily vertify from the redundancy of the stabilizers that the number of the logical qubits indeed equals $k$: the number of stabilizers given equals the number of qubits while the product of these stabilizers over $m$ for a given $j$ is equal to the identity, so there are $j$ redundancies among the stabilizers (and no other redundancies as may easily be seen).

Note of course that in general $U X_{m,j} X_{m+l,j} U^\dagger$ for $j\in {\cal A}$ is a stabilizer of the code, for any $l$, and similarly for $j\in {\cal B}$ with $X$ replaced with $Z$.  This will be useful later.

{\emph For the rest of this paper, we focus on the case $k=2$.}  We will assume ${\cal A}$ is the set $\{1\}$ while ${\cal B}$ is the set $\{2\}$.
In this case, one interesting transversal (up to permutation of qubits) logical operation always exists.  If we interchange qubits $(m,1)$ and $(-m,2)$, then the corresponding interchange of vertices gives an automorphism of $G$.  To see that this is an automorphism, note that if there is an edge between $(m,1)$ and $(m',2)$ with $m'=m+l$, then, by cyclic invariance, there is an edge between $(-m',1)$ and $(-m,2)$ as then $-m=-m'+l$.  Hence, if we apply this permutation of qubits and follow with a Hadamard on all qubits this is an automorphism of the CSS quantum code.
This automorphism of the code implements a combined Hadamard-SWAP operation on the logical qubits; that is, it swaps the two logical qubits and applies a Hadamard to the logical qubits.

A useful case to consider is the case $n_0$ is odd.  In this case, for each $j$, the products $\prod_m Z_{m,j}$ and $\prod_{m} X_{m,j}$ give a pair of anticommuting logical operators.

Another interesting logical operator that must exist is $U^\dagger X_{m,j} U$  for $j\in {\cal A}$ and
$U^\dagger Z_{m,j} U$ for $j\in {\cal B}$.
The existence of this logical gives some bound on the distance of the code: the distance is at most one greater than the minimum degree of a vertex in the graph $G$.

A useful alternative way to label the qubits in the case $k=2$ is by integers $m$ taken mod $n$, with even integers corresponding to the $A$ sublattice and odd corresponding to the $B$ sublattice.  {\emph We will use this method of labeling qubits for the rest of this paper.}
A useful way of writing the graph $G$ in this case is by giving a set ${\cal S}$ of odd integers (modulo $n$) such that there is an edge from vertex $m$ to vertex $m'$, with $m$ even and $m'$ odd, iff $(m'-m) \in S$.  In this case, if $|{\cal S}|$ is even, then we prepare the logical $|+0\rangle$ state, while of $|{\cal S}|$ is odd, we prepare a logical Bell pair.

\section{Rotated Toric Code}
\label{rtc}
We now show that a certain family of rotated toric codes may be written as BCC codes.
A rotated two-dimensional toric code\cite{aasen2025geometrically} may be defined by specifying an integral lattice $\Lambda$ so that the vertices are at coordinates $\mathbb{Z}^2/\Lambda$.  Choosing this lattice $\Lambda$ to be generated by the rows of the matrix
$$
\begin{pmatrix} (d+1)/2 & (d-1)/2  \\ -(d-1)/2 & (d+1)/2 \end{pmatrix}$$
gives a code with distance $d$ for odd $d$\cite{kovalev2013quantum,aasen2025geometrically}.
This lattice has determinant $(d^2+1)/2$ and so this gives a $[[d^2+1,2,d]]$ quantum code.

It is convenient to consider another lattice, the lattice of edges.  Here we consider all edges, both horizontal and vertical, in a single lattice, with the lattice points being taken at the center of the edges.
This lattice is rotated by 45 degrees with respect to the lattice of vertices.  Taking $v_1=(1/2,1/2)$ and $v_2=(1/2,-1/2)$ as basis vectors for translations in this lattice of edges, then translation in the lattice of vertices by $(1,0)$ or $(0,1)$ is implemented by translating by $v_1\pm v_2$, respectively,
We can thus regard the lattice of edges also as an integral lattice: an edge has some new coordinates $(i,j)$, with $i,j$ both integers, if the center of the edge is located at $i v_1 + j v_2$ in the old coordinates.
Since we have modded out the lattice of vertices by some lattice $\Lambda$, similarly the lattice of edges is $\mathbb{Z}^2$ modulo some other lattice $\Lambda_{\rm edge}$.
After some algebra, we may write the Hermite normal form for the matrix whose row vectors generate
$\Lambda_{\rm edge}$ as
$$
\begin{pmatrix} 1 & d  \\ 0 & d^2+1\end{pmatrix},$$
up to some sign changes.

In this way, we are labelling qubits by integers mod $d^2+1$, with even qubits corresponding to horizontal edges and odd qubits corresponding to vertical edges.  Then, $Z$ stabilizers may be chosen to be of the form
$Z_m Z_{m+1} Z_{m+d} Z_{m+d+1}$ for even $m$, for one choice of convention (there are several possible conventions about which stabilizers are $Z$ and which are $X$, and which columns of the Hermite normal form correspond to vertical and which to horizontal).

This state can be obtained by obtained by a BCC code picking ${\cal S}$ to be the set $\{-d,+d,-(d+(d-1)),+(d+(d-1)),-(d+2*(d-1)),+(d+2*(d-1)),\ldots,d^2+1\}$.  For example, for $d=3$ this gives ${\cal S}=\{-3,+3,5\}$, for $d=5$ this gives ${\cal S}=\{-5,+5,-9,+9,13\}$, for $d=7$ this gives ${\cal S}=\{-7,+7,-13,+13,-19,+19,25\}$.
To see this, consider conjugating $Z_{m+1} Z_{m+d}$ by $U$, and note that there is a large cancellation: even though $|{\cal S}|=d$, the resulting stabilizer still only has weight $4$.

In addition to the combined Hadamard-SWAP automorphism that all BCC codes have, this code also has an automorphism that implements logical Hadamard without the swap, in this case given by mapping qubits $m$ to qubit $md$ and implementing a Hadamard on every qubit.

Note that since the given ${\cal S}$ is odd, we prepare a logical Bell pair with this circuit.  There are choices of $S$ which prepare the logical $+0$ state.  For example, if we replace $S$ with the its complement in the set of odd integers $\{1,3,\ldots,n-1\}$ this gives such a set.

{\it Remark---} We may also define a non-CSS code.  We term this a cyclic cluster code, as it is not bipartite.  Such cyclic cluster codes are defined in the same way as BCC codes, except dropping the requirement that the graph $G$ be bipartite.  One such cyclic cluster code is related to the BCC codes considered in this subsection.  Given a set ${\cal S}$ as defined in this section, define a new set ${\cal T}$ by adding $n/2$ to each element of ${\cal S}$, and dividing by $2 \mod n$ (taking the division by $2$ so that the result is even).  For example, for $d=3$, we have ${\cal S}=\{-3,+3,5\}$ so 
${\cal T}=\{-(3+5 \mod n)/2,(+3+5 \mod n)/2,(5+5 \mod 10)/2\}=\{-4,+4,0\}$.  Then define a code by initializing $n/2$ qubits to the $+$ state, and applying CZ gates between qubits $i,j$ if $i-j \mod n/2 \in {\cal T}$.  We drop the element $0$ from ${\cal T}$ if present.  Then for $d=3$ this gives the stabilizers of the non-CSS $[[5,1,3]]$ code.  For higher $d$, we get other elements of the family of non-CSS codes in Example 3 of  \cite{kovalev2013quantum}.

\section{Other Examples}
\label{oe}
We have conducted a numerical search for other BCC codes with $k=2$.  There is an $[[18,2,5]]$ code with $|{\cal S}|=4$ found by searching over choices of ${\cal S}$ and testing the distance for each one.
No $[[16,2,5]]$ BCC code can exist because the smallest CSS code\cite{ezerman2013css} with distance $5$ has $n=17$.
One choice of ${\cal S}$ giving this $[[18,2,5]]$ is ${\cal S}=\{5,11,15,17\}$.
The smallest $[[n,2,7]]$ code with $n/2$ odd and $|{\cal S}|=6$ that we have found has $n=34$, with one possible choice of 
${\cal S}$ being ${\cal S}=\{1,5,7,9,15,23\}$.

The $[[18,2,5]]$ code above enjoys some special properties.
First, in addition to the combined Hadamard-SWAP operation on logical qubits, it also admits a Hadamard operation (without SWAP) on the logical qubits, similar to the rotated toric code.  We discovered the existence of this numerically as follows.  Given a logical operator $O$, let $w_A(O)$ denote the weight of $O$ on the $A$ sublattice, i.e., given that $O$ is a product of Pauli operators, it is the number of Paulis in the product supported on sites in the $A$ sublattice, and let $w_B(O)$ denote the weight of $O$ on the $B$ sublattice.  When searching over logical operators of weight $5$, we found $X$-type logicals with $w_A=1$ and $w_B=4$; such operators must exist by construction for any BCC code.  However, we also found $X$-type logicals with $w_B=1$ and $w_A=4$.  The existence of $18/2=9$ such logicals means that we can disentangle the code state to a state with all qubits on the $B$ sublattice in the $+$ (or $-$) state and all qubits in the $A$ sublattice in the $0$ (or $1$) state using a unitary which is a product of CNOT gates with each qubit in the $B$ sublattice being the source for four such gates.  On investigating the structure of these cyclically invariant gates, we realized that it was given by the same cyclically invariant graph as was used to define the code.  So, in fact the code is self-orthogonal, so that the stabilizer group is invariant under a transversal Hadamard.

Second, this code has some weight $4$ stabilizers.  Note that ${\cal S}$ contains the subset $\{5,11,17\}$ which is invariant under adding $6 \mod 18$.  Thus, when computing the stabilizer $U X_0 X_6 U^\dagger$, many of the Paulis cancel and all that is left is a weight $4$ stabilizer, $X_0 X_6 X_{15} X_{3}$.  Indeed, this may be written more nicely as $X_{-3} X_0 X_3 X_6$, and we have also all shifts of this stabilizer by an even integer.
Indeed, we also have the stabilizer $Z_{-3} Z_0 Z_3 Z_6$ and all shifts of that stabilizer by an even integer.  This may be understood as follows: for each $r\in \{0,1,2\}$, and each $s\in\{0,1,2\}$ we have stabilizers
$X_{2r+6s-3} X_{2r+6s} X_{2r+6(s+1)-3} X_{2r+6(s+1)}$, as well as the same stabilizer with $X$ and $Z$ interchanged.  For each of the three choices of $r$, this gives a quantum code on the $6$ qubits numbered $2r-3,2r,2r+3,2r+6,2r+9,2r+12$, with stabilizers $XXXXII,IIXXXX,XXIIXX$ and the same with $X$ and $Z$ interchanged.  This is a length $3$ repetition code for the operators $XX$ and $ZZ$.
Considering only these weight $4$ stabilizers, we obtain a code on $18$ qubits with $6$ logical qubits and distance $2$; the other, higher-weight stabilizers give the $[[18,2,5]]$ code.  The reader may easily explicitly compute those stabilizers in terms of the logical operators of this $6$ qubit code.

\section{Fault Tolerant Preparation}
\label{ftp}
Finally, we consider fault tolerant circuits to prepare these codes.  We consider only codes of distance $3,5$ in detail.  We consider a simple experiment in which we prepare two copies of the code, and then perform a transversal CNOT gate between them (of course, in some hardware, CZ is the more natural gate and CNOT may be implemented by CZ and Hadamard, and in this case some of the Hadamards we have done to prepare the state cancel against the Hadamards used to implement the CNOT gate with a CZ gate).
We assume that the codes enjoy the property that we may prepare logical $|+0\rangle$ in one copy and logical $|0+\rangle$ in the other, and then we perform a CNOT from the first copy to the second so the result is a nontrivial Bell pair on the first of the two logical qubits.  Finally, we measure this Bell pair in either $X$ or $Z$ basis.  Of course, this is only a partial analysis of fault tolerance, rather than considering its full use in a more complicated circuit.

The naive circuit is simply to prepare the code state as described in \cref{gd}, then apply the transversal gates and then measure properties of the Bell pair.  For this discussion of fault tolerance, we will use $CZ$ gates and assume that both $A$ and $B$ sublattices are prepared in the $+$ state, and then follow with a Hadamard to prepare $+0$, rather than assuming that $A$ is prepared in $+$ and $B$ is prepared in $0$ and using CNOT gates.
The issue that may arise is that if one of the qubits in the $A$ (or $B$) sublattice flips at some point in the circuit when some number $f$ of CZ gates remain on that qubit, then it will produce phase errors on $f$ qubits in the $B$ (respectively, $A$) sublattice, giving a total error of weight $f+1$ from a single error.  While this error has a bit flip error on one sublattice and $f$ phase errors on the other, after Hadamard it becomes either $f+1$ bit flips or $f+1$ phase errors.  This can hurt fault tolerance.  

Note, however, that since flipping a qubit \emph{before} applying any CZ gates in $U$ (i.e., on the initial state) leaves the state invariant, we can replace this error of weight $f+1$ with an error of weight $|{\cal S}|-f$.
So, in the case of distance $3$ codes with $|{\cal S}|=2$, we can always assume that this produces an error of weight $1$ and so fault tolerance is preserved in this case, i.e., the error rate in the Bell pair decays quadratically in the physical error rate.

Similarly, in the case of distance $5$ codes with $|{\cal S}|=4$, a single error can only produce a weight two error.  So, to preserve fault tolerance, it suffices to have some method of checking for a single bit flip.  An easy way to do this is to add $36$ ancilla qubits, one for each data qubit.  Then we consider the following circuit: initialize all data qubits to $+$, copy the state of each data qubit in the $Z$ basis to the corresponding ancilla by initializing the ancilla in $0$ and performing a CNOT (or by initializing the ancilla in $+$ and performing a CZ), implement unitary $U$, then again CNOT from each data to the corresponding ancilla and measure the ancilla in $Z$ (or, do a CZ on data and ancilla and measure the ancilla in $X$).  Then, a single bit flip error will be detected by the ancilla being in $1$ rather than $0$.

This can in fact be done with many fewer ancilla qubits.  We can {\emph share} ancilla between the two code blocks, so that for each $m$, $0\leq m <18$, there is one ancilla.  Then, initialize the ancillas in $0$, perform a CNOT from the $m$-th qubit in each code block to the $m$-th ancilla, for all $m$, perform unitary $U$, then again
perform a CNOT from the $m$-th qubit in each code block to the $m$-th ancilla, and measure all ancilla in the $Z$ basis.  This will again detect any single bit flip error; it can fail to detect if the $m$-th qubit is flipped in both code blocks, which may produce a weight $2$ $Z$ error and a weight $2$ $X$ error, but this still preserves fault tolerance.  Thus, we can perform this with only $36$ data qubits and $18$ ancillas.

We can in fact further reduce the number of ancillas to only $9$, as follows.  A first try to do this is to label the ancilla by $m$, $0\leq m<9$, and let qubits $m,m+9$ in each code all share the same ancilla (so there are four data qubits per ancilla).  We could again CNOT all four of these to the ancilla.  The trouble is that a single bit flip on the ancilla qubit can have a higher weight backaction on the data qubits due to the gates between data and ancilla qubits.  It seems numerically that this may be resolved as follows: prepare the $9$ ancilla in the $+$ state, entangle the ancilla by performing CZ between them (for example, CZ between ancilla $m$ and ancilla $m+3 \mod 9$ for each $m$), then CZ from data to ancilla, perform $U$, CZ from data to ancilla, unentangle the ancilla, and measure them.  This entangled state of the ancilla (which itself is a cyclic cluster state) will detect a single bit ancilla bit flip.
Numerical simulations indicate that these circuits lead to a logical error rate decaying cubically in the physical error rate, i.e., preserving fault tolerance to distance $5$.  For those simulations, samples were preselected which passed all detection criteria using ancilla, and then, after reading out in $Z$ or $X$ basis, a decoder was used which determined, for each block of the code, the minimum number of single qubit errors to produce the observed syndrome.

\bibliography{cyclic-ref}

\begin{thebibliography}{4}
\expandafter\ifx\csname natexlab\endcsname\relax\def\natexlab#1{#1}\fi
\expandafter\ifx\csname bibnamefont\endcsname\relax
  \def\bibnamefont#1{#1}\fi
\expandafter\ifx\csname bibfnamefont\endcsname\relax
  \def\bibfnamefont#1{#1}\fi
\expandafter\ifx\csname citenamefont\endcsname\relax
  \def\citenamefont#1{#1}\fi
\expandafter\ifx\csname url\endcsname\relax
  \def\url#1{\texttt{#1}}\fi
\expandafter\ifx\csname urlprefix\endcsname\relax\def\urlprefix{URL }\fi
\providecommand{\bibinfo}[2]{#2}
\providecommand{\eprint}[2][]{\url{#2}}

\bibitem[{\citenamefont{Kovalev and Pryadko}(2013)}]{kovalev2013quantum}
\bibinfo{author}{\bibfnamefont{A.~A.} \bibnamefont{Kovalev}} \bibnamefont{and}
  \bibinfo{author}{\bibfnamefont{L.~P.} \bibnamefont{Pryadko}},
  \bibinfo{journal}{Physical Review A—Atomic, Molecular, and Optical Physics}
  \textbf{\bibinfo{volume}{88}}, \bibinfo{pages}{012311}
  (\bibinfo{year}{2013}).

\bibitem[{\citenamefont{Briegel and Raussendorf}(2001)}]{briegel2001persistent}
\bibinfo{author}{\bibfnamefont{H.~J.} \bibnamefont{Briegel}} \bibnamefont{and}
  \bibinfo{author}{\bibfnamefont{R.}~\bibnamefont{Raussendorf}},
  \bibinfo{journal}{Physical Review Letters} \textbf{\bibinfo{volume}{86}},
  \bibinfo{pages}{910} (\bibinfo{year}{2001}).

\bibitem[{\citenamefont{Aasen et~al.}(2025)\citenamefont{Aasen, Haah, Hastings,
  and Wang}}]{aasen2025geometrically}
\bibinfo{author}{\bibfnamefont{D.}~\bibnamefont{Aasen}},
  \bibinfo{author}{\bibfnamefont{J.}~\bibnamefont{Haah}},
  \bibinfo{author}{\bibfnamefont{M.~B.} \bibnamefont{Hastings}},
  \bibnamefont{and} \bibinfo{author}{\bibfnamefont{Z.}~\bibnamefont{Wang}},
  \bibinfo{journal}{arXiv preprint arXiv:2505.10403}  (\bibinfo{year}{2025}).

\bibitem[{\citenamefont{Ezerman et~al.}(2013)\citenamefont{Ezerman, Jitman,
  Ling, and Pasechnik}}]{ezerman2013css}
\bibinfo{author}{\bibfnamefont{M.~F.} \bibnamefont{Ezerman}},
  \bibinfo{author}{\bibfnamefont{S.}~\bibnamefont{Jitman}},
  \bibinfo{author}{\bibfnamefont{S.}~\bibnamefont{Ling}}, \bibnamefont{and}
  \bibinfo{author}{\bibfnamefont{D.~V.} \bibnamefont{Pasechnik}},
  \bibinfo{journal}{IEEE Transactions on Information Theory}
  \textbf{\bibinfo{volume}{59}}, \bibinfo{pages}{6732} (\bibinfo{year}{2013}).

\end{thebibliography}
\end{document}